\documentclass{emulateapj}

\usepackage{caption}

\shorttitle{Discovery of gamma-ray emission from the SNR RCW~86 with H.E.S.S.}
\shortauthors{F.~A. Aharonian et al. (H.E.S.S. Collaboration)}

\begin{document}

\sloppy


\title{Discovery of gamma-ray emission from the shell-type supernova remnant RCW 86 with H.E.S.S.}


\author{F. Aharonian~\altaffilmark{1,13},
  A.G.~Akhperjanian~\altaffilmark{2},
  U.~Barres de Almeida~\altaffilmark{8,a},
  A.R.~Bazer-Bachi~\altaffilmark{3},
  B.~Behera~\altaffilmark{14},
  M.~Beilicke~\altaffilmark{4},
  W.~Benbow~\altaffilmark{1},
  K.~Bernl\"ohr~\altaffilmark{1,5},
  C.~Boisson~\altaffilmark{6},
  A.~Bochow~\altaffilmark{1}
% ...
}

% ...
\altaffiltext{1}{
Max-Planck-Institut f\"ur Kernphysik, P.O. Box 103980, D 69029
Heidelberg, Germany}
\altaffiltext{2}{
 Yerevan Physics Institute, 2 Alikhanian Brothers St., 375036 Yerevan,
Armenia}
\altaffiltext{3}{
Centre d'Etude Spatiale des Rayonnements, CNRS/UPS, 9 av. du Colonel Roche, BP
4346, F-31029 Toulouse Cedex 4, France}
\altaffiltext{4}{
Universit\"at Hamburg, Institut f\"ur Experimentalphysik, Luruper Chaussee
149, D 22761 Hamburg, Germany}
\altaffiltext{5}{
Institut f\"ur Physik, Humboldt-Universit\"at zu Berlin, Newtonstr. 15,
D 12489 Berlin, Germany}
\altaffiltext{6}{
LUTH, Observatoire de Paris, CNRS, Universit\'e Paris Diderot, 5 Place Jules Janssen, 92190 Meudon, 
France}
% ...
\altaffiltext{8}{
University of Durham, Department of Physics, South Road, Durham DH1 3LE,
U.K.}
% ...
\altaffiltext{13}{
Dublin Institute for Advanced Studies, 5 Merrion Square, Dublin 2,
Ireland}
\altaffiltext{14}{
Landessternwarte, Universit\"at Heidelberg, K\"onigstuhl, D 69117 Heidelberg, Germany}
% ...
\altaffiltext{a}{supported by CAPES Foundation, Ministry of Education of Brazil}
% ...


\begin{abstract}
The shell-type supernova remnant (SNR) RCW 86, possibly associated 
with the historical supernova SN 185, with its relatively large size 
(about 40' in diameter) and the presence of non-thermal X-rays is a 
promising target for $\gamma$-ray observations. The high sensitivity, 
good angular resolution of a few arc minutes and the large field of view 
of the High Energy Stereoscopic System (H.E.S.S.) make it ideally suited for 
the study of the $\gamma$-ray morphology of such extended sources. 
H.E.S.S. observations have indeed led to the 
discovery of the SNR RCW~86 in very high energy (VHE; $\rm{E} > 100$~GeV) 
$\gamma$-rays. With 31 hours of observation time, the source is 
detected with a statistical significance of $8.5 \sigma$ and 
is significantly more extended than the H.E.S.S. point spread function. Morphological 
studies have been performed and show that the $\gamma$-ray flux does not correlate 
perfectly with the X-ray emission. The flux from the 
remnant is $\sim$10\% of the flux from the Crab nebula, with a similar 
photon index of about 2.5. Possible origins of the very high 
energy gamma-ray emission, via either Inverse Compton scattering by electrons or 
the decay of neutral pions produced by proton interactions, 
are discussed on the basis of spectral features obtained both in the X-ray and $\gamma$-ray regimes.
\end{abstract}





\sloppy

\section{Introduction}
Shell-type supernova remnants (SNR) are widely believed to be the prime 
candidates for accelerating cosmic ray protons and nuclei up to $10^{15}$eV. A promising 
way of proving the existence of high energy hadrons 
accelerated in SNR shells is the detection of very high energy (VHE; E $> 100$~GeV) 
$\gamma$-rays produced in nucleonic interactions with ambient matter. 
VHE $\gamma$-ray emission has been detected recently in several shell-type SNRs, especially 
from Cassiopeia~A~(\cite{casa1}, \cite{casa}), RX~J1713.7-3946~\citep{HESSRXJ3} and RX~J0852.0-4622~\citep{HESSVelaJr}. These two latest sources both show an extended morphology highly correlated 
with the structures seen in non-thermal X-rays. Although a hadronic origin is 
probable in the above cases \citep{rxjberezhko}, a leptonic origin can not be ruled out \citep{porter}.\\ 
Another young shell-type SNR is RCW~86 (also known as 
G315.4-2.3 and MSH~$14-6{\it 3}$). It has a complete shell in 
radio~\citep{kesteven}, optical~\citep{smith} and 
X-rays~\citep{pisarski}, with a nearly circular shape of $40$' 
diameter. It received substantial attention because of its possible 
association with SN~185, the first historical Galactic 
supernova~\citep{clarck}. However, conclusive evidence for this connection is still 
missing: using optical observations,~\cite{rosado} found 
an apparent kinematic distance of 2.8~kpc and an age of $\sim10\,000$ years, 
whereas recent observations of the North-East part of the remnant with 
the Chandra and XMM-Newton satellites strengthen the case that the 
event recorded by the Chinese in 185~AD was a supernova and that RCW 86 is its 
remnant~\citep{vink}. In this case, a distance to the SNR of $\sim1$~kpc 
can be estimated for a standard Sedov evolution scenario~\citep{bocchino}. 
The X-ray spectrum obtained with the Einstein satellite was first 
represented by a two-temperature 
plasma model~\citep{winkler}. Then, RXTE~\citep{petre} and ASCA observations 
(\cite{bamba}, \cite{borkowski}), with a wider spectral coverage, were used to resolve 
a non-thermal component in the X-ray spectrum which can be well described by a 
soft power-law with a photon index of $\sim3$. 
The large-scale density gradient across RCW~86 (\cite{pisarski} and \cite{claas}) possibly 
suggests that the northern part could be the shocked half of a very low-density wind bubble 
plus dense shell from the progenitor star, and to this extent it could well be similar to 
RX~J1713.7-3946 and RX~J0852.0-4622. In its southern part, RCW~86 contains an HII region. Apparently, 
the gas density in this HII region is rather high and spatially extended. Therefore, the SNR shock
has swept over an extended high density region in the South, with consequent high radio
and thermal X-ray emissions~\citep{bocchino}. 
With a diameter of about 40', RCW~86 is one of the very few non-thermal X-ray emitting 
SNRs resolvable in VHE $\gamma$-rays. H.E.S.S., with its high sensitivity, its good angular resolution 
and its large field of view is ideally 
suited for morphology studies of such an extended object.\\
Evidence for $\gamma$-ray emission from RCW 86 was found using the 
CANGAROO-II instrument, but no firm detection was claimed~\citep{cangaroo}. 
Here, we present data on RCW~86 obtained 
with the full H.E.S.S. array between 2004 and 2007.

\section{H.E.S.S. observations and analysis methods}
H.E.S.S. is an array of four imaging Cherenkov telescopes located 
1800~m above sea level in the Khomas Highland in 
Namibia~\citep{HESS}. Each telescope has a tesselated mirror with an 
area of $107 \, \rm{m^{2}}$~\citep{HESSOptics} and is equipped with a camera 
comprising 960 photomultipliers~\citep{HESSCamera} covering a field of 
view of 5$^{\circ}$ in diameter. Due to the effective rejection of 
hadronic air showers with the stereoscopic imaging technique, the 
H.E.S.S. telescope system can detect point sources near zenith at 
flux levels of about 1\% of the Crab nebula flux with a 
statistical significance of 5~$\sigma$ standard deviation in 25 hours of observation
~\citep{HESSCrab}. \\
The shell-type SNR RCW~86 was observed between 2004 and 2007 with the 
complete H.E.S.S. array. After standard data quality selection and dead time 
correction, the resulting live time is 31 hours. The observations have 
been carried out at zenith angles ranging from $38^{\circ}$ 
to $53^{\circ}$. The data were taken using the wobble mode where the source is offset from the centre 
of the field of view, alternating between 28 minute runs in the 
positive and negative declination or right ascension directions; 
the mean offset angle of the data set used in this analysis is $0.7^{\circ}$. 
The energy threshold of the system increases with zenith angle: for the observations presented 
here, the average threshold was 480 GeV.\\
The data were calibrated using standard H.E.S.S. calibration 
procedures, as discussed by \cite{HESSCalib}. The data were analyzed 
using a Hillas parameter based method as described in \cite{HESSHillas} with 
\emph{standard} cuts, which include a minimum requirement of 80 photo 
electrons in each camera image. Two different background estimation 
procedures were used, as described in \cite{Davidbackground}. For 2D image 
generation and morphology studies, the \emph{ring background} method was 
applied with a mean ring radius of $0.7^{\circ}$. As this method uses an energy 
averaged radial acceptance correction, the \emph{reflected-region background} 
method was applied for spectral studies. In 
this second background subtraction procedure, OFF events were selected from 
the same field of view and in the same runs as the ON events by 
selecting the region symmetric to the ON region with respect to the 
camera centre. As a cross-check, a second analysis chain, sharing 
only the raw data and using the ``Combined Model'' analysis \citep{denaurois}, 
was also applied to the data. The two analysis methods yield consistent results.

\section{Results} 
\begin{figure}[ht]
\plotone{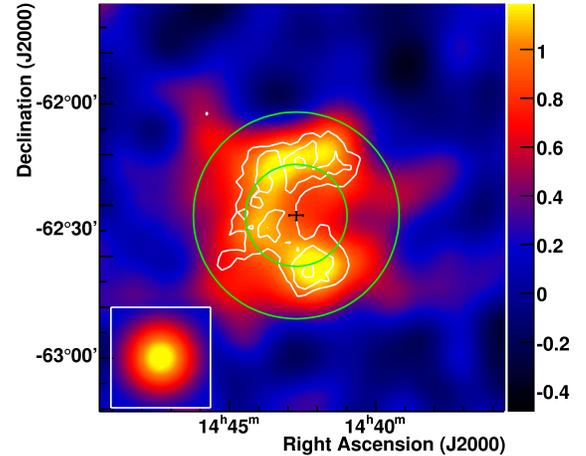}
\caption{H.E.S.S. $\gamma$-ray image of RCW~86. The map was smoothed with a Gaussian function with a $\sigma_{\rm{smooth}}=4.8'$ to reduce the effect of statistical fluctuations. The linear color scale is in units of excess counts per arcmin$^{2}$. White contours correspond to 4, 5, 6 $\sigma$ significance, 
obtained by counting gamma rays within 0.14$^{\circ}$ from each given location. The image inset in the bottom left 
corner indicates the size of a point source as seen by H.E.S.S., for an equivalent analysis, smoothing and zenith angles. The centre of 
the fitted shell, as discussed in the text, is marked by a black cross. The two solid green circles correspond to the inner and outer radii of this shell. 
\label{fig::rcw86}}
\end{figure}

A clear VHE $\gamma$-ray signal of $8.5 \sigma$ standard deviation and $1546 \pm 183$ 
excess $\gamma$-rays is detected from a circular region of $0.45^{\circ}$ 
radius, centered on ($\alpha_{J2000}$ = 14$^h$42$^m$43$^s$, 
$\delta_{J2000}$ = $-62^\circ28'48''$). This integration region was chosen 
a priori on the basis of the X-ray data obtained with the ROSAT satellite and fully 
encompasses the SNR. Figure~\ref{fig::rcw86} shows the VHE 
$\gamma$-ray excess map of the $1.6^{\circ} \times 1.6^{\circ}$ region around RCW~86. 
The map has been smoothed with a Gaussian kernel with a $\sigma$ of $4.8'$ to 
suppress statistical fluctuations on scales smaller than the 
H.E.S.S. point-spread function (PSF). 
The VHE $\gamma$-ray excess from RCW~86 is significantly extended beyond the 
PSF of the instrument, which is illustrated in the bottom left corner of Figure~\ref{fig::rcw86}. 
Contours of constant significance are superimposed in white at the 4, 5 and 6$\sigma$ levels. 
An excess map has also been produced with the so-called ``hard cuts'' for better gamma hadron separation, 
which includes a stricter cut of 200 photo electrons on the image size compared to the ``standard cuts'', 
and was found to be compatible with Figure~\ref{fig::rcw86}. The VHE emission shown in Figure~\ref{fig::rcw86} 
is suggestive of a shell-like morphology. To test this hypothesis, the brightness profile of a thick shell 
projected along the line of sight and folded with the H.E.S.S. point-spread 
function was fit to the unsmoothed excess map. As illustrated in Figure~\ref{fig::rcw86}, 
the best fit ($\chi^2/\rm{ndf}$ = 233.1/220) is obtained with an outer radius of $24.43' \pm 1.79'_{\rm stat}$, 
a width of $12.39' \pm 4.22'_{\rm stat}$ 
and a centre of the shell at ($\alpha_{J2000}$ = 14$^h$42$^m$42.96$^s \pm 14.1^s_{\rm stat}$, 
$\delta_{J2000}$ = $-62^{\circ}26'41.6'' \pm 66.5''_{\rm stat}$). 

\begin{figure}[ht]
\plotone{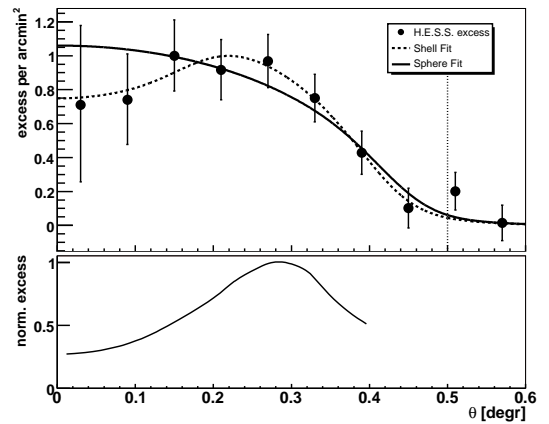}
\caption{\textbf{Upper panel:} H.E.S.S. radial profiles around the fitted centre of the SNR ($\alpha_{J2000}$ = 14$^h$42$^m$43$^s$,
$\delta_{J2000}$ = $-62^\circ$26'42''). The solid line shows the result of a projected uniformly-emitting sphere smoothed with the H.E.S.S. point-spread function and fitted to the H.E.S.S. data. The dashed line corresponds to a projection of a thick and spherically symmetric shell. The dotted vertical line illustrates the extent of the region used for the azimuthal profile and for the spectral analysis. \textbf{Lower panel:} Radial profiles of the X-ray data (3-6 keV) from XMM-Newton. These data are background subtracted and smoothed to match the H.E.S.S. angular resolution. Additionally, the obtained excess profile was normalized.} 
\label{fig::radprof}
\end{figure}

\begin{figure}[ht]
\plotone{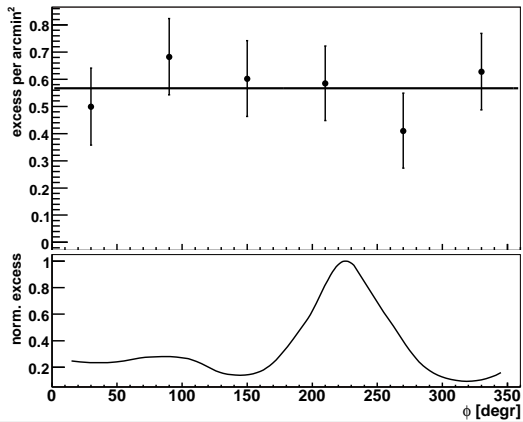}
\caption{\textbf{Upper panel:} H.E.S.S. azimuthal profile integrated over a region of $0.5^{\circ}$ radius covering the SNR RCW~86. The azimuthal angle is calculated with respect to the fitted shell centre. $0^{\circ}$ corresponds to the North part of the source and $90^{\circ}$ to the East. The solid line shows the result of a fit of the data to a constant which yields a chi-square of 1.47 for 5 degrees of freedom. \textbf{Lower panel:} Azimuthal profiles of the X-ray data (3-6 keV) from XMM-Newton. These data are background subtracted and smoothed to match the H.E.S.S. angular resolution. Additionally, the obtained excess profile was normalized.} 
\label{fig::azprof}
\end{figure}

Figure~\ref{fig::radprof} shows the radial profile of the VHE excess 
relative to the fitted centre. The fit of the radial profiles to the data points results in a 
chi-square per degree of freedom of $\chi^2/\rm{ndf}$ = 2.85/7 for a projected shell (determined by outer ring radius,
ring width and absolute normalization) which is not significantly 
better than the fit of a projected uniformly-emitting sphere characterized by a ring radius and
a normalization factor ($\chi^2/\rm{ndf}$ = 5.43/8). 
Also visible in Figure~\ref{fig::rcw86} is an apparent deficit of $\gamma$-rays at the western part of the SNR. 
However, the azimuthal profile in Figure~\ref{fig::azprof} is consistent with a constant and reveals 
that this dip is not significant ($\chi^2/\rm{ndf}$ = 1.47/5).\\

\begin{figure}[htbp]
\plotone{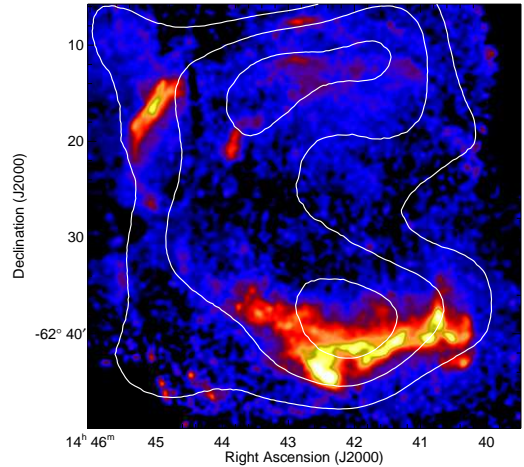}
\caption{Excess contours of $\gamma$-ray emission (0.55, 0.8, 1.05 $\gamma$-rays per arcmin$^{2}$ Gaussian smoothed with $\sigma_{\rm{smooth}}=4.8'$) superimposed on the background subtracted XMM-Newton EPIC (MOS/PN) 3-6 keV X-ray image of the remnant. 
\label{fig::xmm}}
\end{figure}

Figure~\ref{fig::xmm} shows the 3-6 keV X-ray map of RCW~86 obtained using six observations of the remnant 
carried out by the XMM-Newton satellite in 2006~\citep{vink} and additional observations taken in 2007. 
The energy range was selected to avoid as much as possible contamination from line emission from the, in general, 
cool plasma ($<$ 1 keV) of RCW 86. Potentially, the 3 - 4 keV range could contain some contamination from Ar and Ca 
lines, but no such line emission is seen in the available Chandra, XMM-Newton~\citep{vink} or Suzaku 
spectra~\citep{ueno}. This map was obtained by first automatically cleaning the observations of 
$> 3 \sigma$ excursions to the mean count rate, thus minimizing the background of the maps.
Then, for each observation and for each of the three detectors (MOS1, MOS2, and PN), a background count rate in 
the 3-6 keV band was determined using a relatively empty region of the field of view. 
In the final stage, the background image was subtracted from the count rate image, and then corrected using the 
exposure maps obtained with the standard XMM-Newton SAS 7.1.0 software (which includes vignetting correction), 
in order to obtain the background corrected map displayed in Figure~\ref{fig::xmm}. 
An overall positional agreement with the H.E.S.S. contours derived from 
Figure~\ref{fig::rcw86} as well as a good compatibility between the outer radius of the 
$\gamma$-ray emission ($24.43' \pm 1.79'_{\rm stat}$) and the extension of the X-ray emission can be observed. 
However, the emission peak apparent in the X-ray azimuthal profile is not visible in $\gamma$-rays 
(Figure~\ref{fig::azprof}). Furthermore, the dip in surface brightness at 
the center of the remnant seems more pronounced in the X-ray radial profiles (Figure~\ref{fig::radprof}). 
A more detailed comparison of the 
$\gamma$-ray and X-ray morphologies would require higher statistics than presently available, and hence 
will have to await future longer observations.\\

\begin{figure}[htbp]
\plotone{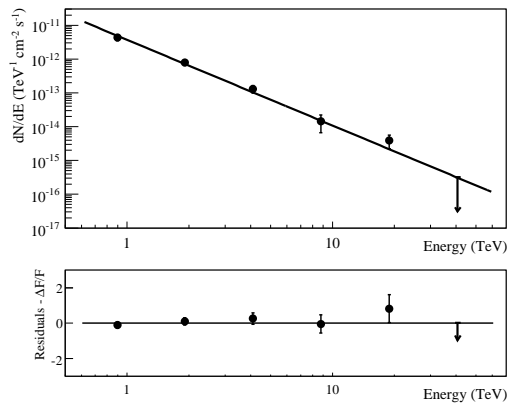}
\caption{Differential energy spectrum of RCW~86, extracted from a
circular region of $0.5^{\circ}$ radius around the position ($\alpha_{J2000}$ = 14$^h$42$^m$43$^s$,
$\delta_{J2000}$ = $-62^\circ$26'42'') adjusted to the H.E.S.S. data 
to enclose the whole source. The solid line shows the result of a pure power-law fit. The error bars denote 1$\sigma$ statistical
errors; the upper limit (arrow) is estimated at the $2\sigma$ level. The bottom panel shows the residuals to the power-law fit. Events with energies between 600 GeV and 60 TeV were used in the determination of the spectrum.
\label{fig::spec}}
\end{figure}

For the spectral analysis, the source region (ON region) is defined by 
a circle of $0.5^{\circ}$ radius centered on the best fit 
position of the shell, chosen to fully enclose the whole source. The 
radius of the extraction region is illustrated in Figure~\ref{fig::radprof}. 
The spectrum obtained (see Figure~\ref{fig::spec}) is well described 
by a power-law with a photon index of $2.54 \pm 0.12_{\rm{stat}} \pm 0.20_{\rm{sys}}$ and a 
flux normalisation at 1 TeV of $(3.72 \pm
0.50_{\rm{stat}} \pm 0.8_{\rm{sys}}) \times 10^{-12} \mathrm{cm^{-2}} \mathrm{s^{-1}}
\mathrm{TeV^{-1}}$ ($\chi^2/\rm{ndf}$ = 6.30/4). The integral flux in the energy range 1 - 10 TeV 
is $(2.34 \pm 0.3_{\rm{stat}} \pm 0.5_{\rm{sys}}) \times 10^{-12} \, \mathrm{cm^{-2}} \mathrm{s^{-1}}$, 
which corresponds to $\sim$ 10\% of the integrated flux of the Crab nebula in 
the same energy interval. No significant improvement is obtained by fitting a power-law with an exponential 
cut-off ($\chi^2/\rm{ndf}$ = 2.96/3). If the fit range is 
restricted to energies below 10 TeV, a photon index of $2.41 \pm 0.16_{\rm{stat}} \pm 0.20_{\rm{sys}}$ and a flux 
normalisation at 1 TeV of $(3.57 \pm 0.5_{\rm{stat}} \pm 0.8_{\rm{sys}}) \times 10^{-12} 
\mathrm{cm^{-2}} \mathrm{s^{-1}} \mathrm{TeV^{-1}}$ are determined ($\chi^2/\rm{ndf}$ = 0.68/2), compatible with the fit of the SNR in the whole energy range. \\

\newpage

\section{Discussion}
There are two commonly invoked mechanisms for VHE $\gamma$-ray production in young 
supernova remnants, inverse Compton (IC) scattering of high energy 
electrons off ambient photons (leptonic scenario) and $\pi^0$ meson 
production in inelastic interactions of accelerated protons with ambient 
gas (hadronic scenario). In such a hadronic scenario, a comparison between the expected thermal X-ray 
emission and the actually measured thermal emission has to await deeper observations in which one can better 
determined whether the TeV emission traces the denser, thermal X-ray emitting parts of the SNR, or is more 
closely correlated with the X-ray synchrotron emission from the remnant.\\ 
The measured $\gamma$-ray spectrum from RCW~86, restricted to energies below 10 TeV, 
translates into an energy flux between 1 and 10 TeV of $8.6 \times 10^{-12} \, \rm erg \, cm^{-2} \, 
s^{-1}$. The X-ray spectrum of the whole remnant is mixed between thermal and non-thermal emission. 
Assuming that the hard X-ray continuum originates from non-thermal synchrotron emission as reported by~\cite{rho}, 
\cite{vink} and \cite{ueno}, the measurement made by~\cite{petre} using RXTE data provides an estimate of the 
total amount of non-thermal flux from RCW~86. They find that the spectrum is well fitted by a power-law of
index $\sim3$ and a flux normalization at 10 keV of $10^{-4} \, \rm cm^{-2} \, s^{-1} \, keV^{-1}$, 
which extrapolated down to the 0.7 to 10 keV band leads to an integral flux of 
$2.1 \times 10^{-10} \, \rm erg \, cm^{-2} \, s^{-1}$. 
In a leptonic scenario, assuming that the $\gamma$-ray emission 
is entirely due to the IC process on cosmic microwave background photons, 
the ratio of the synchrotron power and IC power radiated  
is often used to constrain the magnetic field.  
For a power-law distribution of electron energies, $K\gamma^{-p}$, the general equation relating the 
synchrotron power ($P_S$) produced by electrons with Lorentz factors between $\gamma_{1,X}$ and $\gamma_{2,X}$ 
and the IC power ($P_{IC}$) radiated between $\gamma_{1,IC}$ and $\gamma_{2,IC}$ can be expressed as follow:
\begin{eqnarray}
\frac{P_S}{P_{IC}} = \frac{U_B}{U_{ph}} \frac{(\gamma_{2,X}^{3-p} - \gamma_{1,X}^{3-p})}{(\gamma_{2,IC}^{3-p} 
- \gamma_{1,IC}^{3-p})}
\label{eq1}
\end{eqnarray}
where $U_{ph}$ and $U_B$ are the energy density of the photon field and the energy 
density of the magnetic field, respectively. It should be noted here that, for a fixed X-ray energy,
 $\gamma_{1,X}$ and $\gamma_{2,X}$ are inversely proportional to the square root of the magnetic field.
If X-rays and $\gamma$-rays probe the same region of the electron spectrum, one finds the 
standard relation between the synchrotron and IC power $\frac{P_{S}}{P_{IC}} = \frac{U_B}{U_{ph}}$. 
Assuming that the target photon field is the cosmic microwave background,
a magnetic field of $30 \, \mu \rm G$ can be estimated using Equation~\ref{eq1} and the 
synchrotron photon index of $\sim 3$, 
independent of the distance and age of the SNR. This estimate is compatible with that of~\cite{vink} 
based on thin filaments resolved by Chandra (assuming a distance of 2.5~kpc) in which the authors also 
deduce a high speed of the blast wave ($\sim 2700 \, \rm km \, s^{-1}$); their estimated value would 
increase to $\sim 50 \, \mu \rm G$ for a distance of 1~kpc. However, it is
still a factor of 2 lower than the maximum field strength determined by~\cite{volk} using a lower 
shock velocity of $800 \, \rm km \, s^{-1}$ as suggested by
optical data in the Southern region of the SNR~\citep{rosado}. The difference between the 
field amplification estimated by~\cite{vink} and that of~\cite{volk} lies in the fact that V\"olk et al. 
obtained a higher result when they de-projected the measured filament width, as for an ideal spherical shock, 
whereas Vink et al. did not. Without de-projection the two results remarkably agree, even though they were 
obtained for the southern side and the northern side, respectively. A 
discussion of de-projection for RCW~86 is given in~\cite{volk}. With similar data,~\cite{bamba2} 
deduced a significantly lower magnetic field strength of $\sim 4-12 \, \mu \rm{G}$. However, their 
analysis is based on rather different assumptions on the nature of filament formation.\\ 
In a hadronic scenario, one can estimate the total energy in accelerated
protons $W_p$ in the range $10 - 100$~TeV required to produce the
$\gamma$-ray luminosity $L_{\gamma}$ observed by H.E.S.S. using the
relation $W_p (10 - 100 \, \rm{TeV}) \approx \tau_{\gamma} \times L_{\gamma}(1
- 10 \, \rm{TeV})$, in which $\tau_{\gamma} \approx 4.9 \times 10^{15} \left(\frac{n}{1 \, 
\mathrm{cm^{-3}}} \right)^{-1} \, \rm{s}$ is the characteristic 
cooling time of protons through the $\pi^0$ production channel~\citep{kelner}. 
The total energy injected in protons is calculated by
extrapolating the proton spectrum down to 1 GeV. 
Because of this extrapolation over 4 decades in energy, the uncertainty of the estimate can be as 
large as a factor of 10. Assuming that the relatively steep slope of the proton spectrum (as
inferred from the observed $\gamma$-ray spectrum) 
is the result of an energy cut-off (somewhere around several tens of TeV in proton energy), and that at
lower energies the proton spectrum has a $E^{-2}$ type spectrum representative of those predicted
by the diffusive shock acceleration theory, the total energy budget in all protons for
the distance of 2.5 kpc and the ambient gas density between $0.3 \, \rm cm^{-3}$ and
$0.7 \, \rm cm^{-3}$~\citep{bocchino}, would be $(2-4) \times 10^{50} \, \rm
erg$. This estimate is in reasonable agreement with theoretical expectations that a
significant fraction of the explosion energy of $10^{51} \, \rm erg$ is released in
relativistic protons. On the other hand, if the power-law spectrum of protons continues
to GeV energies with the spectral index $\Gamma=2.4$ (i.e. similar to the gamma-ray spectrum below 10 TeV), 
the total budget in protons would
exceed a few times $10^{51}$ erg for a distance of 2.5~kpc. This would exclude the hadronic origin of TeV $\gamma$-rays, 
unless the SNR is nearby ($\sim 1$ kpc), or the $\gamma$-rays are produced in very dense regions. 
Indeed, \cite{pisarski} and \cite{claas} reported that there is a large density contrast across the remnant, 
e.g. in the South, where the density could be as high as $10 \, \rm cm^{-3}$; with such a dense medium, a larger distance 
for the remnant could still be compatible with the observed $\gamma$-ray flux.

\section{Conclusions}
H.E.S.S. observations have led to the discovery of the shell-type SNR
RCW~86 in VHE $\gamma$-rays. The $\gamma$-ray signal is significantly more extended 
than the H.E.S.S. point-spread function. The possibility of a
shell-like morphology was addressed, but cannot be settled on the
basis of the limited statistics available at the moment. The flux from
the remnant is $\sim$10\% of that from the Crab nebula, with a 
photon index of about 2.5. The question of 
the nature of the particles producing the $\gamma$-ray signal observed 
by H.E.S.S. is also discussed.\\ 
In a leptonic scenario, assuming that the 
$\gamma$-ray emission is entirely due to the IC process on cosmic microwave 
background photons and that the synchrotron and IC photons are produced by 
the same electrons, the ratio of the $\gamma$-ray energy flux and the 
X-ray flux determines the magnetic field to be close to $30 \, \mu \rm G$.\\
In the hadronic scenario, the lack of information about the low-energy $\gamma$-ray spectrum 
results in large uncertainties on the total energy budget in protons. If below several tens of TeV, 
the proton spectrum has a $E^{-2}$ type spectrum, the total energy
in protons would be in reasonable agreement with theoretical expectations.
On the other hand, if we assume that the proton spectrum continues down
to GeV energies with the observed spectral index $\Gamma = 2.4$, energetics
would rule out a hadronic origin for the TeV $\gamma$-rays unless the SNR
is nearby, or if the $\gamma$-rays are produced in a very dense medium as
reported in the southern part of the remnant.

\acknowledgments
The support of the Namibian authorities and of the University of Namibia
in facilitating the construction and operation of H.E.S.S. is gratefully
acknowledged, as is the support by the German Ministry for Education and
Research (BMBF), the Max Planck Society, the French Ministry for Research,
the CNRS-IN2P3 and the Astroparticle Interdisciplinary Programme of the
CNRS, the U.K. Science and Technology Facilities Council (STFC),
the IPNP of the Charles University, the Polish Ministry of Science and 
Higher Education, the South African Department of
Science and Technology and National Research Foundation, and by the
University of Namibia. We appreciate the excellent work of the technical
support staff in Berlin, Durham, Hamburg, Heidelberg, Palaiseau, Paris,
Saclay, and in Namibia in the construction and operation of the
equipment.


\begin{thebibliography}{}
\bibitem[Aharonian et al.(2001)]{casa1} Aharonian, F., 2001, A\&A, 112, 307
\bibitem[Aharonian et al.(2004)]{HESSCalib} Aharonian, F., 
({\it H.E.S.S. Collaboration}) 2004, APh, 22, 109
\bibitem[Aharonian et al.(2005)]{HESSHillas} Aharonian, F., et al.
({\it H.E.S.S. Collaboration}) 2005,  A\&A, 430, 865
\bibitem[Aharonian et al.(2006)]{HESSCrab} Aharonian, F., et al.
({\it H.E.S.S. Collaboration}) 2006, A\&A, 457, 899
\bibitem[Aharonian et al.(2007a)]{HESSRXJ3} Aharonian, F., et al.
({\it H.E.S.S. Collaboration}) 2007a, A\&A, 464, 235
\bibitem[Aharonian et al.(2007b)]{HESSVelaJr} Aharonian, F., et al.
({\it H.E.S.S. Collaboration}) 2007b, A\&A, 661, 236
\bibitem[Albert et al.(2007)]{casa} Albert, J., et al. 2007, A\&A, 474, 937
\bibitem[Bamba et al.(2000)]{bamba} Bamba, A., Koyama, K., \& Tomida, H., 2000, PASJ, 52, 1157
\bibitem[Bamba et al.(2005)]{bamba2} Bamba, A., Yamazaki, R., Yoshida, T., Terasawa, T., \& Koyama, K., 2005, 
ApJ, 621, 793
\bibitem[Berge et al.(2007)]{Davidbackground} Berge, D., Funk, S., \& Hinton, J.,
 2007, A\&A, 466, 1219
\bibitem[e.g. Berezhko \& V\"olk (2006)]{rxjberezhko} Berezhko, E. G., \& V\"olk, H. J., 2006, A\&A, 451, 981
\bibitem[Bernl\"ohr et al.(2003)]{HESSOptics} Bernl\"ohr, K., et al., 2003, APh, 20, 111
\bibitem[Bocchino et al.(2000)]{bocchino} Bocchino, F., Vink, J., Favata, F., Maggio, A., \& Sciortino, S., 2000, 
A\&A, 360, 671
\bibitem[Borkowski et al.(2001)]{borkowski} Borkowski, K. J., Arnaud, K. A., Dorman, B., Hughes, J. P., Sarazin, C. L., \& Smith, R. A., 2001, ApJ, 550, 334
\bibitem[Claas et al.(1989)]{claas} Claas, J.J., Kaastra, J. S., Smith, A., Peacock, A., \& de Korte, P. A. J., 1989, ApJ, 337, 399
\bibitem[Clark \& Stephenson (1977)]{clarck} Clark, D., \& Stephenson, F., 1977, The Historical Supernovae (Oxford: Pergamon Press), 83
\bibitem[Hinton (2004)]{HESS} Hinton, J. A., 2004, NewAR, 48, 331
\bibitem[Kelner et al.(2006)]{kelner} Kelner, S. R., Aharonian, F. A., \& Bugayov, V. V., 2006, Physical Review D, 74, 3
\bibitem[Kesteven \& Caswell (1987)]{kesteven} Kesteven, M. J., \& Caswell, J. L., 1987,  A\&A, 183, 118
\bibitem[de Naurois et al.(2005)]{denaurois} de Naurois, M. et al. 2005, 
 in Proceedings of the conference ``Towards a Network of Atmospheric Cherenkov Detectors VII'', ed. B. Degrange \& G. Fontaine (Palaiseau: Ecole Polytechnique), 173
\bibitem[Petre et al.(1999)]{petre} Petre, R., Allen, G. E., \& Hwang, U.,  1999, 
Astron. Nachr., 320, 199
\bibitem[Pisarski et al.(1984)]{pisarski} Pisarski, P. L., Helfand, D. J., \& Kahn, S. M., 1984, 
ApJ, 277, 710
\bibitem[e.g. Porter et al.(2006)]{porter} Porter, T. A., Moskalenko, I. V., \& Strong, A. W. 2006, ApJ, 648, L29
\bibitem[Rho et al.(2002)]{rho} Rho, J., Dyer, K. K., Borkowski, K. J., \& Reynolds, S. P., 2002, ApJ, 581, 1116 
\bibitem[Rosado et al.(1996)]{rosado} Rosado, M., Ambrocio-Cruz, P., Le Coarer, E., \& Marcelin, M., 1996, 
A\&A, 315, 243
\bibitem[Smith (1997)]{smith} Smith, R. C., 1997, 
AJ, 114, 2664
\bibitem[Ueno et al.(2007)]{ueno} Ueno, M., et al. 2007, PASJ, 59, 171
\bibitem[Vincent et al.(2003)]{HESSCamera} Vincent, P., et al. 2003, in
  Proceedings of the 28th International Cosmic Ray Conference,
  T. Kajita et al., Eds. (Universal Academy Press, Tokyo, 2003), 2887
\bibitem[Vink et al.(2006)]{vink} Vink, J., Bleeker, J., Van Der Heyden, K., Bykov, A., Bamba, A., \& Yamazaki, R., 2006, 
ApJL, 648, 33
\bibitem[V\"olk et al.(2005)]{volk} V\"olk, H. J., Berezhko, E. G., \& Ksenofontov, L. T., 2005, 
A\&A, 433, 229
\bibitem[Watanabe et al.(2003)]{cangaroo} Watanabe, S., et al. ({\it CANGAROO Collaboration}) 2003, in
  Proceedings of the 28th International Cosmic Ray Conference, IUPAP, Eds: T. Kajita et al., 2397
\bibitem[Winkler (1978)]{winkler} Winkler Jr., P. F., 1978, ApJ, 221, 220
\end{thebibliography}
\end{document}